\documentclass[preprint,showpacs,preprintnumbers,amsmath,amssymb,superscriptaddress,10pt,graphicx]{revtex4}

\usepackage{graphicx}
\usepackage{dcolumn}
\usepackage{bm}
\usepackage{amssymb}

\begin{document}

\title{Quantum Yang-Mills Condensate Dark Energy Models}

\author{Wen Zhao}
\affiliation {School of Physics and Astronomy, Cardiff University, Cardiff, CF24 3AA, United Kingdom}

\author{Yang Zhang}
\affiliation {Center for Astrophysics, University of Science and Technology of China, Hefei, 230026, People's Republic of China}

\author{Minglei Tong}
\affiliation {Center for Astrophysics, University of Science and Technology of China, Hefei, 230026, People's Republic of China}


\date{\today}


\begin{abstract}

We review the quantum Yang-Mills condensate (YMC) dark energy models. As the
effective Yang-Mills Lagrangian is completely determined by the quantum field
theory, there is no adjustable parameter in the model except the energy scale. In
this model, the equation-of-state (EOS) of the YMC dark energy, $w_y > -1$ and
$w_y < -1$, can both be naturally realized. By studying the evolution of various
components in the model, we find that, in the early stage of the universe, dark
energy tracked the evolution of the radiation, i.e. $w_y \rightarrow 1/3$.
However, in the late stage, $w_y$ naturally runs to the critical state with $w_y =
-1$, and the universe transits from matter-dominated into dark energy dominated
stage only at recently $z \sim 0.3$. These characters are independent of the
choice of the initial condition, and the cosmic coincidence problem is avoided in
the models. We also find that, if the possible interaction between YMC and dust
matter is considered, the late time attractor solution may exist. In this case,
the EOS of YMC must evolve from $w_y>0$ into $w_y < -1$, which is slightly
suggested by the observations. At the same time, the total EOS in the attractor
solution is $w_{tot} = -1$, the universe being the de Sitter expansion in the late
stage, and the cosmic big rip is naturally avoided. These features are all
independent of the interacting forms.

\end{abstract}



\maketitle


\section{Introduction}

The observations of Type Ia Supernova (SNIa) \cite{sn}, together with the cosmic microwave background radiation (CMB)\cite{map}
and the larger scale structure \cite{sdss}, suggest that the present universe is accelerating expansion, which needs a kind of mysterious dark energy
with negative equation-of-state (EOS). The simplest model is by introducing the cosmological constant term $\Lambda$, which has a constant
effective  EOS $w=-1$, and drive the acceleration of the universe. If assuming the effective energy of the $\Lambda$ term occupies $\sim 73\%$
of the total energy, together with $\sim 23\%$ dark matter, $\sim 4\%$ baryon matter and $\sim 10^{-5}$ radiation component, constitute the so-called
$\Lambda$CDM model.

Although this simple model satisfies nearly all the cosmological observations, it still remains a phenomenological  model. Since the major components,
$\Lambda$ and dark matter, still keep unclear for us. For the $\Lambda$ as the candidate of dark energy, also suffers from the following dilemmas (see for instant \cite{constant}):
First, the effective energy density is quite tiny, $\rho_{\Lambda}\sim 5.8 h^2\times 10^{-11}$eV$^4$. If we consider it as the vacuum energy of the particle physics,
this energy density is $120$ order smaller than the Planck energy scale! This is the so-called `fine-tunning' problem. Second, the density scale of $\Lambda$
keeps constant in all the stage of the universe. The observations show that
the present value of the matter component (including dark matter and baryon components) is
some one third of $\rho_{\Lambda}$,
but it  varies with time as $\rho_m \propto a^{-3}$.
So,  for example,  at an earlier time of radiation-matter equality
with  redshift $z \sim 3454$,
$\rho_{\Lambda}$ should be a very fine
tuned value $\sim 6.3\times 10^{-11} \rho_m$.
Otherwise,
a slightly variant initial value of $\rho_{\Lambda}$ would lead to
a value of the ratio  $\rho_{\Lambda}/\rho_m$
drastically different from the observed one.
This  is the so-called `coincidence' problem.
In addition, the $\Lambda$CDM model also faces some observational challenges: there are mild evidences show that, the EOS of dark energy might evolve from $w>-1$ in
the early stage to $w<-1$ in the current stage \cite{cross2}, which is expected to be confirmed by the future sensitive observations \cite{detf}.

So, it is necessary to look for other candidates as the dark energy, especially the dynamical models.  One possibility is proposing a canonical scalar field $\phi$ with the langrangian $\mathcal{L}_{\phi}=\dot{\phi}^2/2-V(\phi)$, which is dubbed as the `quintessence' models (see the review \cite{scalarreview}). Similar to the inflationary field, when the potential term $V(\phi)$ is dominant, the EOS of the quintessence field approaches to $-1$, i.e. $\Lambda$-like, and accounts for the observations. The most interesting is that, in \cite{track-quint}, the authors introduced a kind of potential forms, such as $V(\phi)=M^{4+\alpha}\phi^{-\alpha }$ or $V(\phi)=M^4[\exp(M_{\rm pl}/\phi)-1]$, which have the tracker solutions, i.e. the field $\phi$ tracked the evolution of the background components in the early universe. So they address the coincidence problem, i.e. removing the need to tune initial conditions in order for the matter and dark energy  densities to nearly coincide today. Although, this kind of models have been excluded by the cosmological observations, it provides the excellent possibility to naturally avoided the coincidence problems.

However, the quintessence models also suffer from some dilemmas. The EOS of the quintessence models are in the range $-1<w<1$. In order to obtain a EOS with $w<-1$, one always has to introduce the so-called `phantom' field, which includes the non-canonical negative kinetic terms \cite{phantom}\cite{quintom}. However, the `phantom' field lacks the strong physical motivation, and also leads to the problem of quantum instabilities \cite{stable}. In \cite{bigrip}, the authors also pointed that, if $w<-1$ keeps, the universe shall face the `big-rip' problem.

In addition to proposing the dynamical dark energy models, some efforts have been paid to modify the general relativity (GR) \cite{gravity}, which can also speed-up the universe in the recent stage. However, it should be pointed that, any revised GR should prepare to go through the strict solar system test \cite{solar}, as well as to explain the various cosmological observations, such as the CMB temperature and polarization anisotropies power spectra. In addition, to our view, the current observations have not provided the strong reasons to answer: Why we should modify GR and how to modify it.

In this paper, we shall propose the Yang-Mills condensate (YMC) dark energy models, where, instead of the scalar field, the quantum Yang-Mills field is considered as the candidate of dark energy component. Recently, the similar models are also discussed by a number of authors \cite{vector}\cite{emfield}. Different from the scalar field models, the Yang-Mills fields are the indispensable cornerstone to particle physics, gauge bosons have been observed. There is no room for adjusting the form of effective Yang-Mills Lagrangian as it is predicted by quantum corrections according to field theory. In this review, we shall firstly introduce physical motivations of the YMC dark energy models in Section II.  In Section III, we simply introduce and discuss the Lagrangian of effective quantum Yang-Mills field.

As the main part of this paper, in Section IV, we apply the YMC into the cosmology as the candidate of dark energy, and investigate the cosmic evolution of the various components, especially the evolution of dark energy. We find the excellent characters of this kind of models.  Different from the quintessence models, both the EOS $w_y>-1$ and $w_y<-1$ can be naturally realized. In the free YMC dark energy models, $w_y\rightarrow 1/3$ in the early stage and tracked the evolution of radiation component. Only in the recent stage, $w_y$ rolls to the  critical state with $w_y=-1$, i.e. $\Lambda$-like, and accounts for the observations. This feature is independent of the choice of the initial condition, so the coincidence problem is naturally avoided. We also find that, if the possible interaction between YMC and dust components is considered, the late time attractor solution can exist, where the EOS of YMC naturally runs from $w_y\rightarrow1/3$ to the the phantom-like attractor state, i.e. $w_y<-1$. However, the total effective EOS is $w_{tot}=-1$ in the attractor solution, and the so-called `big-rip' problem is also avoided. We should pointed that, these new features are all independent of the interaction forms. In Section V, we calculate the statefinder and $Om$diagnostics for the YMC dark energy models, which are helpful to differentiate the YMC models from the other dark energy models from the observations.

Section VI is contributed as a summary of this paper.

Throughout this paper, we will work with unit in which
$c=\hbar =k_B=1$.

\section{physical motivation}

The introduction of the quantum effective YMC
into cosmology \cite{y}  has been motivated  by
the fact that the $SU(3)$ YMC has given a phenomenological description
of the vacuum within hadrons confining quarks,
and yet at the same time all the important properties
of a proper quantum field are kept,
such as the Lorentz invariance,
the gauge symmetry, and the correct trace anomaly \cite{23} \cite{25}.
Quarks inside a hadron  would experience the existence
of the Bag constant,  $B$,
which is equivalent to an energy density $B$ and  a pressure $-B$.
So quarks would feel an energy-momentum  tensor
of the vacuum as $T_{\mu \nu}= B {\rm diag} (1,-1,-1,-1)$.
This non-trivial vacuum has been formed mainly
by the contributions from the quantum effective YMC,
and from the possible interactions with quarks.
Our thinking has been that if the vacuum inside a hadron
is filled with the quantum effective YMC,
what if the vacuum of the universe as a whole is also
filled with some kind of YMC.

Gauge fields play a very important role in,
and are the indispensable cornerstone to,  particle physics.
All known fundamental interactions between particles
are mediated through gauge bosons.
Generally speaking, as a gauge field,
the YMC under consideration may have
interactions with other species of particles in the universe.
However, unlike those well known interactions in QED, QCD,
and the electro-weak unification,
here   at the moment we do not yet have a model
for the details of the microscopic interactions between
the YMC and other particle.
Therefore, in this paper on the dark energy model based on
the quantum effective YMC,
 we will adopt a simple description of the possible interactions
between the YMC and other cosmic particles, in addition to the simplest model with free YMC component.
We will investigate in these models
the cosmic evolution of the universe
from the radiation-dominated era up to the  present.


\section{Yang-Mills field model}

In the effective YMC dark energy model, the effective Yang-Mills field Lagrangian is given by \cite{23}\cite{zzcqg2006,zzplb2006}:
 \begin{eqnarray}\label{l_eff}
 \mathcal{L}_{\rm eff}=\frac{1}{2}bF\left(\ln\left|{F}/{\kappa^2}\right|-1\right),
 \end{eqnarray}
where $\kappa$ is the renormalization scale of dimension of squared mass, $F\equiv -(1/2)F^{a}_{\mu\nu}F^{a\mu\nu}$ plays the role of the order parameter of the YMC. The Callan-Symanzik coefficient $b=(11N-2N_f)/24\pi^2$ for $SU(N)$ with $N_f$ being the number of quark flavors. For the gauge  group $SU(2)$ considered here, one has $b=11/12\pi^2$ when the fermion's contribution is neglected. For the case of $SU(3)$ the effective Lagrangian in (\ref{l_eff}) leads to a phenomenological description of the asymptotic freedom for quarks inside hadrons \cite{23}. It should be noted that the $SU(2)$ Yang-Mills field is introduced here as a model for the cosmic dark energy, in may not be directly identified as the QCD fields, nor the weak-electromagnetic unification gauge fields.

An explanation can be given for the form in (\ref{l_eff}) as an effective Lagrangian up to 1-loop quantum correct \cite{23,24,25}. A classical $SU(N)$ Yang-Mills field Lagrangian is $\mathcal{L}=F/2g_0^2$, where $g_0$ is the bare coupling constant. As is known. when 1-loop quantum corrections are included, the bare coupling $g_0$ will be replaced by a running one $g$ as the following \cite{26}, $g_0^2\rightarrow g^2=2/(b\ln(k^2/k_0^2))$, where $k$ is the momentum transfer and $k_0$ is the energy scale. To build up an effective theory \cite{23,24,25}, one may just replace the momentum transfer $k^2$ by the field strength $F$ in the following manner: $\ln(k^2/k_0^2)\rightarrow 2\ln|F/e\kappa^2|$, yielding equation (\ref{l_eff}).  We would like to point out that the renormalization scale $\kappa$ is the only parameter of this effective Yang-Mills model, and its value should be determined by comparing the observations. In contrast to the scalar field dark energy models, the YMC Lagrangian is completely fixed by quantum corrections up to 1-loops, and there is no room for adjusting its functional form. This is an attractive feature of the effective YMC dark energy model. We should mention that, the YMC dark energy models including 2-loop and 3-loop quantum corrects are also discussed in the recent papers \cite{xzplb2007}\cite{wzxjcap2008}.  It was found that, these more complicated models have the exactly same characters with the 1-loop models. So in this paper, we mainly focus on the 1-loop models with the Lagrangian in (\ref{l_eff}).

\section{YMC as dark energy}

Let us work in the flat Friedmann-Lemaitre-Robertson-Walker (FLRW) universe, which is described by
\begin{eqnarray}\label{metric}\nonumber
 ds^2=dt^2-a^2(t)\delta_{ij}dx^idx^j=a^2(\tau)[d\tau^2-\delta_{ij}dx^idx^j],
 \end{eqnarray}
 where $t$ time and $\tau$ time are related by $cdt=ad\tau$. Considering the simplest case with only the YMC in the universe, which minimally coupled to the gravity,
the effective action is,
 \begin{eqnarray}
 S=\int \sqrt{-{g}}~\left[-\frac{\mathcal R}{16\pi G}+\mathcal{L}_{\rm eff}\right] ~d^{4}x.
 \label{S}
 \end{eqnarray}
Here, ${g}$ is the determinant of the metric $g_{\mu\nu}$. $\mathcal{R}$ is the scalar Ricci curvature, and $\mathcal{L}_{\rm eff}$ is the effective Lagrangian of YMC, described by Eq. (\ref{l_eff}).
By variation of $S$ with respect to the metric $g^{\mu\nu}$, one
obtains the Einstein equation $G_{\mu\nu}=8\pi GT_{\mu\nu}$, where
the energy-momentum tensor of YMC is given by
 \begin{eqnarray}\label{T_munu}
 T_{\mu\nu}=\sum_{a=1}^{3}~\frac{g_{\mu\nu}}{4g^2}F_{\sigma\delta}^a
 F^{a\sigma\delta}+\epsilon F_{\mu\sigma}^aF^{a\sigma}_{\nu}.
 \label{T}
 \end{eqnarray}
The dielectric constant is defined by
$\epsilon\equiv2\partial\mathcal{L}_{\rm eff}/\partial F$. In the one-loop
order, it is given by
 \begin{eqnarray}
 \epsilon=b\ln\left|{F}/{\kappa^2}\right|.\label{epsilon}
 \end{eqnarray}
This energy-momentum tensor (\ref{T_munu}) is the sum of three different
energy-momentum tensors of the vectors,
$T_{\mu\nu}=\sum_a~^{(a)}T_{\mu\nu}$, neither of which is of
prefect-fluid form. Here we assume the gauge fields are only the
functions of time $t$, and
$A_{\mu}=\frac{i}{2}\sigma_aA_{\mu}^a(t)$ (here $\sigma_a$ are the
Pauli's matrices) are given by $A_0=0$ and $A_i^a=\delta_i^aA(t)$. Thus, we will find that,
the total energy-momentum tensor $T_{\mu\nu}$ is
homogeneous and isotropic.

Define the Yang-Mills field tensor as usual:
 \begin{eqnarray}
 F^{a}_{\mu\nu}=\partial_{\mu}A_{\nu}^a-\partial_{\nu}A_{\mu}^a+f^{abc}A_{\mu}^{b}A_{\nu}^{c},
 \end{eqnarray}
where $f^{abc}$ is the structure constant of gauge group and
$f^{abc}=\epsilon^{abc}$ for the $SU(2)$ case. This tensor can be
written in the form with the electric and magnetic field as
 \begin{eqnarray}
 F^{a\mu}_{~~\nu}=\left(
 \begin{array}{cccc}
      0 & E_1 & E_2 & E_3\\
     -E_1 & 0 & B_3 & -B_2\\
     -E_2 & -B_3 & 0 & B_1\\
     -E_3 & B_2 & -B_1 & 0
 \end{array}
 \right).
 \end{eqnarray}
It can be easily found that $E_1^2=E_2^2=E_3^2$, and
$B_1^2=B_2^2=B_3^2$. Thus $F$ has a simple form with $F=E^2-B^2$,
where $E^2=\sum_{i=1}^3E_i^2$ and $B^2=\sum_{i=1}^3B_i^2$. In this
case, each component of the energy-momentum tensor is
 \begin{eqnarray}
 ^{(a)}T_{\mu}^{0}&=&\frac{1}{6}(\epsilon-b)(B^2-E^2)\delta^{0}_{\mu}+\frac{\epsilon}{3}
 E^2\delta^{0}_{\mu},\\
 ^{(a)}T_{j}^{i}&=&\frac{1}{6}(\epsilon-b)(B^2-E^2)\delta^i_j+\frac{\epsilon}{3}E^2\delta^i_j\delta^a_j
 -\frac{\epsilon}{3}B^2\delta^i_j(1-\delta^a_j).
 \end{eqnarray}
Although this tensor is not isotropic, its value along the $j=a$
direction is different from the ones along the directions
perpendicular to it. However, the total energy-momentum
tensor $T_{\mu\nu}=\sum_{a=1}^3~^{(a)}T_{\mu\nu}$ has isotropic
stresses, and the corresponding energy density and pressure are
given by
\begin{eqnarray}
\rho_y=\frac{1}{2}\epsilon(E^2+B^2)+\frac{1}{2}b(E^2-B^2),~~p_y=\frac{1}{6}\epsilon(E^2+B^2)-\frac{1}{2}b(E^2-B^2).
\end{eqnarray}

In this paper, for simplicity, we only discuss the pure `electric' case, $F=E^2$. This is a typical consideration, since in the expanding universe, a given `magnetic' component of Yang-Mills field decreases quite rapidly, and the Yang-Mills field becomes the `electric' type \cite{zxijmpd2007}. The energy density and pressure of YMC are reduced to
   \begin{eqnarray}\nonumber
  \rho_y=\frac{E^2}{2}(\epsilon+b),~~p_y=\frac{E^2}{2}\left(\frac{\epsilon}{3}-b\right).
  \end{eqnarray}
It is convenient to introduce a dimensionless quantity $y\equiv \epsilon/b=\ln|F/\kappa^2|$.  So the quantities $\rho_y$ and $p_y$ can be rewritten as
   \begin{eqnarray}\label{rho_p}
  \rho_y=\frac{1}{2}b\kappa^2(y+1)e^y,~~p_y=\frac{1}{6}b\kappa^2(y-3)e^y.
  \end{eqnarray}
One sees that, to ensure that energy density $\rho_y$ be positive in any physically viable model, the allowance for the quantity $y$ should be $y>-1$, i.e. $F>\kappa^2/e\simeq 0.368\kappa^2$. The EOS of the YMC is
 \begin{eqnarray}\label{eos}
 w_y\equiv\frac{p_y}{\rho_y}=\frac{y-3}{3y+3}.
 \end{eqnarray}
Before setting up a cosmological model, the EOS $w_y$ itself as a function of $F$ is interesting.  At the state with $F=\kappa^2$, which is called critical point, one has $y=0$ and the YMC has an EOS of the cosmological constant with $w_y=-1$. Around this critical point, $F<\kappa^2$ gives $y<0$ and $w<-1$, and $F>\kappa^2$ gives $y>0$ and $w>-1$. So in the YMC dark energy models, EOS with $w>-1$ and $w<-1$ all can be naturally realized. On the other hand, in the high energy scale with $F\gg\kappa^2$, one finds that the YMC exhibits an EOS of radiation with $w_y=1/3$. In the follows, we will detailed show that the YMC was evolving from the state with $w_y=1/3$ to $w_y=-1$ (or even to $w_y<-1$) in the expanding universe. This is the main context in this paper. In addition, the characteristic statefinder parameters and the perturbations of the YMC dark energy is also discussed in \cite{zijmpd2008,tzxijmpd2009}\cite{zraa2009}, which are helpful to distinguish the YMC model from other dark energy models.

As is known, an effective theory is a simple representation for an interacting quantum system of many degrees of freedom at and around its respective low energies. Commonly, it applies only in low energies. However, it is interesting to note that the YMC model as an effective theory intrinsically incorporates the appropriate states for both high and low temperature. As has been shown, the same expression in (\ref{eos}) simultaneously gives $p_y\sim-\rho_y$ at low energies, and $p_y\rightarrow\rho_y/3$ at high energies. Therefore, the model of effective YMC can be used even at higher energies than the renormalization scale $\kappa$.

\subsection{Free YMC models\label{section3.1}}

Let us discuss the cosmological model, which filled with three kinds of major energy components, the dark energy, the matter, including both baryons and dark matter, and the radiation. Here, the dark energy component is represented by YMC, and the matter component is simply described by a non-relativistic dust with negligible pressure, and the radiation component consists of the photons and possibly other particles, such as the neutrino, if they are massless.

Since the universe is assumed to be flat, the sum of the fraction densities is $\Omega_{y}+\Omega_m+\Omega_r=1$, where the fractional energy densities are $\Omega_{y}\equiv \rho_y/\rho_{tot}$,   $\Omega_{m}\equiv \rho_m/\rho_{tot}$,  $\Omega_{r}\equiv \rho_r/\rho_{tot}$, and the total energy density is $\rho_{tot}\equiv\rho_y+\rho_m+\rho_r$.  The overall expansion of the universe is determined by the Friedmann equations:
\begin{eqnarray}\label{friedmann1}
 &&\left(\frac{\dot{a}}{a}\right)^2=\frac{8\pi G}{3}(\rho_y+\rho_m+\rho_r),\\ \label{friedmann2}
 &&\frac{\ddot{a}}{a}=-\frac{4\pi G}{3}(\rho_y+3p_y+\rho_m+\rho_r+3\rho_r),
\end{eqnarray}
 where and what follows, the superscript $dot$ denote the $d/dt$. These three components of energy contribute to the source on the right-hand side of the equations. We should notice that, $p_r=\rho_r/3$. In this subsection, we shall assume there is no interaction between these three components. The dynamical evolutions of the three components are determined by their equations of motion, which can be written as equations of the conservation of energy,
 \begin{eqnarray}\label{a1}
 &&\dot{\rho}_y+3\frac{\dot{a}}{a}(\rho_y+p_y)=0,\\ \label{a2}&&\dot{\rho}_m+3\frac{\dot{a}}{a}\rho_m=0,\\ \label{a3}&&\dot{\rho}_r+3\frac{\dot{a}}{a}(\rho_r+p_r)=0.
 \end{eqnarray}
As is known, Eq. (\ref{friedmann2}) is not independent and can be derived from Eqs. (\ref{a1}), (\ref{a2}), (\ref{a3}) and (\ref{friedmann1}). From Eqs. (\ref{a2}) and (\ref{a3}), we can obtain the evolutions of the matter and radiation components, $\rho_m\propto a^{-3}$ and $\rho_m\propto a^{-4}$.

Now, let us focus on the evolution of YMC. Inserting (\ref{rho_p}) into (\ref{a1}), we can obtain the following simple relation,
 \begin{eqnarray}\label{y_evolution}
 ye^{y/2}\propto a^{-2},
 \end{eqnarray}
where the coefficient of proportionality in the above depends on
the initial condition. At the early stages, $a \rightarrow 0$, Eq. (\ref{y_evolution})
leads to $y\gg1$, and Eq. (\ref{eos}) gives
$w_y\rightarrow1/3$, so the YMC behaves as the
radiation component. With the expansion of the universe, the value of $y$ runs to the critical state of $y=0$, and the EOS goes to $w_y=-1$. So, in the late stage of the universe, the YMC behaves as the cosmological constant.
This is one of the most attractive character of the YMC models. Around the critical point with $y=0$,  Eq.(\ref{y_evolution}) yields
$y\propto a^{-2}$,
and the EOS of YMC is
$w_y+1\simeq{4y}/{3}\propto a^{-2}$. The YMC can achieve the states of $w_y>-1$ and $w_y<-1$,
but it can not cross over $-1$, just like in the scalar models \cite{cross}.

 \begin{figure}
 \centerline{\includegraphics[width=18cm,height=10cm]{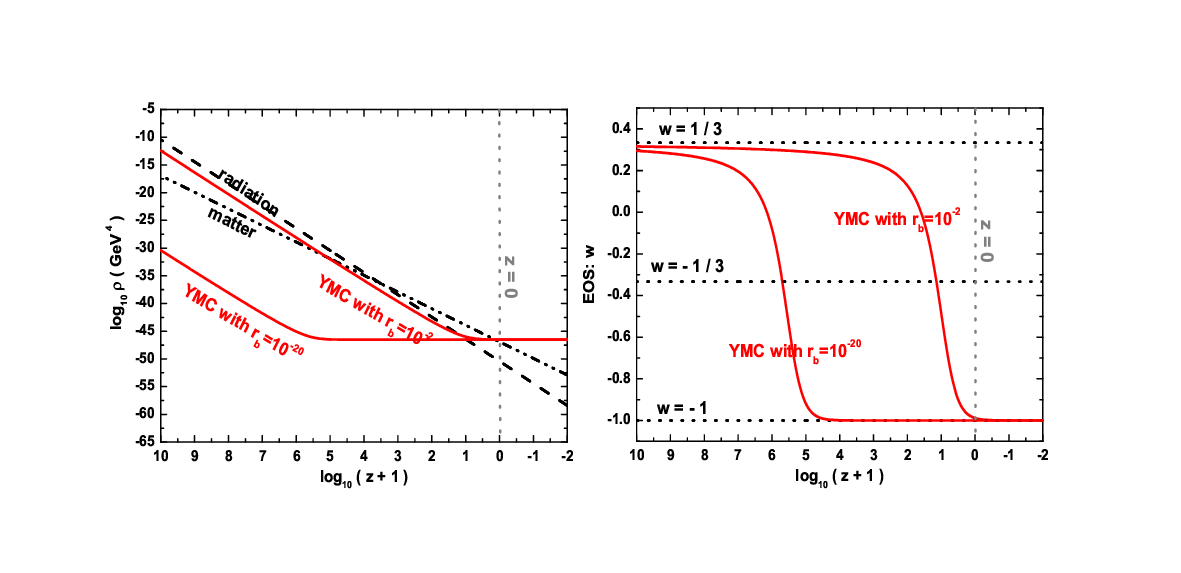}}
 \caption{ In the free YMC dark energy models, the evolution of the YMC energy density (left panels) and EOS (right panels) for the models with different initial conditions. \label{fig1}}
 \end{figure}

 \begin{figure}
 \centerline{\includegraphics[width=18cm,height=10cm]{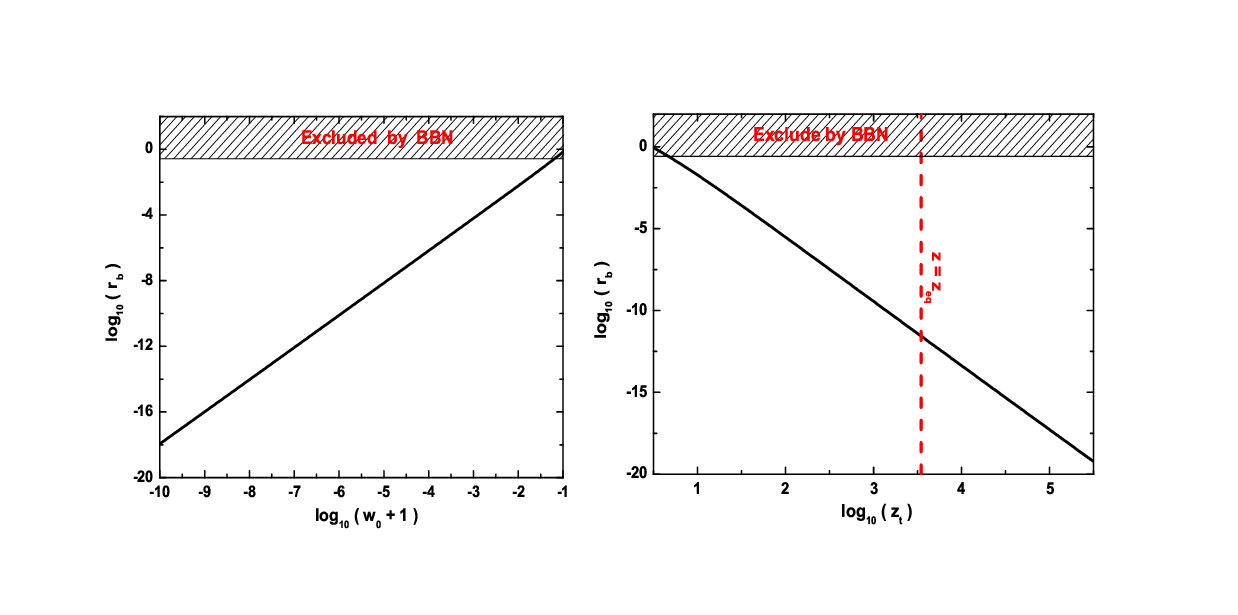}}
 \caption{ Left panel: The present EOS of the YMC $w_0$ depends on the initial value $r_b$. Right panel: The transition redshift $z_t$ depends on the initial value $r_b$.
 In both panels, the shadow region has been excluded by the observation of BBN.  \label{fig2}}
 \end{figure}

We should notice that, the relation in (\ref{y_evolution}) can also be derived by the effective Yang-Mills equations.  By variation of the action $S$ with respect to gauge field $A_{\mu}^{a}$, we can obtain the Yang-Mills equations, which are
 \begin{eqnarray} \label{ym_equation}
 \partial_{\mu}(a^4\epsilon~
 F^{a\mu\nu})+f^{abc}A_{\mu}^{b}(a^4\epsilon~F^{c\mu\nu})=0.
 \label{F1}
 \end{eqnarray}
The $\nu=0$ component of which is an identity, and the $\nu=1,2,3
$ spatial components can be reduced to
 $\partial_{\tau}(a^2\epsilon E)=0$.
At the critical point ($\epsilon=0$), this equation is an
identity. When $\epsilon\neq 0$, this equation has an exact
solution as (\ref{y_evolution}) \cite{zzplb2006}.

Now, let us fix the value of $\kappa$, the only parameter in the
model. At the present time, the YMC energy density is
 \begin{eqnarray}
 \rho_y=\frac{bE^2}{2}(y+1)\simeq\frac{b\kappa^2}{2},
 \end{eqnarray}
and, as the dark energy, it should be $ \Omega_{y}\rho_{tot}$, where
the present total energy density in the universe $\rho_{tot}\approx
8.099 \,h^2\times 10^{-11}\rm {eV}^4$. We choose $\Omega_{y}=0.73$ as
has been observed,  yielding
 \begin{eqnarray}  \label{kappa}
 \kappa=3.57 \, h\times 10^{-5} {\rm eV}^2.
 \end{eqnarray}
This  energy scale is low compared to typical energy scales in
particle physics, such as the QCD and the weak-electromagnetic unification. This is the reason, why the $SU(2)$ Yang-Millls field introduced in this paper cannot
be directly identified as the the QCD gluon fields, nor the weak-electromagnetic unification gauge fields.

To be more specific about how the YMC evolves in the
expanding universe, we look at an early stage when the Big Bang
nucleosynthesis (BBN) processes occur around a redshift $z\sim
10^{10}$ with an energy scale $\sim 1$MeV. To see how the
evolution of $\rho_y$  depends the the initial condition, we
introduce the ratio of energies of the two components
 \begin{eqnarray}
 r_b\equiv\left.\frac{\rho_y}{\rho_r}\right|_{z=10^{10}},
 \end{eqnarray}
where  $\rho_r$ is the radiation energy density. We consider
$r_b<1$, i.e. the YMC is subdominant to the radiation
component initially. Of course, the YMC evolves
differently for different initial values of $r_b$. Nevertheless,
we will see that, as the result of evolution, the present universe
is always dominated by the YMC $\Omega_y \sim 0.73$ for
a very wide range of initial values $r_b$.

Now we use the exact solution (\ref{y_evolution}) to plot the evolution of
$\rho_y$  as a function of the redshift $z$ in Fig. \ref{fig1} (left panel). As
specific examples, here we take  $r_b=10^{-2}$, and
$r_b=10^{-20}$. In comparison, also plotted are the energy
densities of radiation, and of matter. It is seen that,  in the
early stages,  $\rho_y$  decreases as $\rho_y\propto a^{-4}$. So
the YMC density is subdominant and tracks the radiation, a scaling
solution. The corresponding EOS of Yang-Mills field is $w_y \simeq1/3$
shown in Fig. \ref{fig1} (right panel). At late stages,  with the expansion of the
universe, $a\rightarrow \infty$,
 $y$ decreases to nearly zero, and
$w_y\rightarrow-1$ asymptotically. Moreover,  this asymptotic
region is arrived at some redshift $z$ before the present time,
and this $z$ has different values for different initial values of
$r_b$. For smaller $r_b$, the transition redshift is larger (seen
in Fig. \ref{fig1}), and the transition happens earlier. Once the
asymptotic region is achieved, the density of the YMC levels
off and remain a constant forever, like a cosmological constant.
We have also checked that the present value $\Omega_y \sim 0.73$
is also the outcome of the cosmic evolution for any value of $r_b$
in the very wide range $(10^{-20}, 10^{-2})$. So the coincidence
problem do not exist in the YMC model.

The present EOS of the YMC $w_0$  is nearly $-1$. Fig. \ref{fig2} (left panel) plots the
dependence of the present EOS $w_0$ on the initial condition
$r_b$. The  function $\log_{10}(r_b)$ versus
$\log_{10}(w_0+1)$ is nearly linear: a smaller $r_b$ leads to
a smaller $w_0$. For a value $r_b=10^{-2}$, one has
$w_{0}=-0.99$. For a value $r_b=10^{-20}$, $w_{0}$ would
be $-1$ accurately  up to one in $10^{11}$. Therefore, at present
the YMC  is very similar to the cosmological constant.

The solution in Eq.(\ref{y_evolution}) can converted into the following
form
 \begin{eqnarray}
 z=\sqrt{\frac{y}{y_0}}\exp\left[\frac{y-y_0}{4}\right]-1,
 \end{eqnarray}
where $y_0$ is the value of $y$ at $z=0$, depending on the
initial value $r_b$. For a fixed $y_0$, this formula tells a
one-one relation between the EOS (through $y$) and the
corresponding redshift $z$. As is seen from Fig. \ref{fig1}, the
transition of $\omega$ from $1/3$ to $-1$ occurs during a finite
period of time, instead of instantly. To characterize the time of
transition, we use $z_t$ to denote the redshift when
$\omega=-1/3$, i.e. $y=1$,  as given by Eq.(\ref{eos}). This
is,  in fact, the time when the strong energy condition begins to
be violated, i.e., $\rho_y+3 p_y \leq 0$. Then
 \begin{eqnarray}
 z_t=\sqrt{\frac{1}{y_0}}\exp\left[\frac{1-y_0}{4}\right]-1.
 \end{eqnarray}
Therefore, this gives a function $z_t=z_t(r_b)$. Fig. \ref{fig2} (right panel) shows
how the transition redshift $z_t$ depends on the ratio $r_b$.
Interestingly, this transition can occur before, or after the
radiation-matter equality ($z_{eq}=3454$). This feature
is different from the tracked quintessence models in
which transition occurs during the matter dominated era
\cite{track-quint}. A larger $r_b$ leads to a smaller
$z_t$. For example, $r_b=10^{-2}$ leads to  $z_t\simeq 12.4\ll
z_{eq}$, and the transition occurs in the matter dominated stage,
and $r_b=10^{-20}$ leads to $z_t \simeq 5.0\times10^{5}\gg
z_{eq}$, and the transition occurs in the radiation dominated
stage.

The  value of  $r_b$ cannot be chosen to arbitrarily large. In
fact, there is a constraint from the observation result of the
BBN. As is known, the presence of dark energy during
nucleosynthesis epoch will speedup the expansion, enhancing  the
effective species $N_{\nu}$ of neutrinos \cite{bbn}.
 The latest analysis gives a constrain on
 the  extra neutrino species $\delta N_{\nu}\equiv N_{\nu}-3 <1.60$ \cite{bbn}.
Here in our model, the dark energy is played by the YM field. By a
similar analysis, the ratio $r_b$ is related to $\delta N_{\nu}$
through $  r_b= \frac{7\delta  N_{\nu}/4}{10.75} $. This leads to
an upper limit $r_b<0.26$, the present EoS $w_{0}<-0.94$ by
Fig. \ref{fig2} (left panel), and the transition redshift $z_t>5.8$ by Fig. \ref{fig2} (right panel). The
range of initial conditions $r_b\in (10^{-20}, 10^{-2})$ that we
have taken satisfies this constraint.

\subsection{Coupled YMC models\label{section3.2}}

In this subsection, we shall generalize the original YMC dark
energy model to include the interaction between the YMC and dust
matter. We should mention that, the possible interaction between
YMC and radiation component may also exist, which has been
discussed in \cite{xzzcqg2007}. (The similar models on the scalar
field dark energy have been discussed by a number of authors (see
\cite{wangbing} for instant)).  In this section, we assume the YMC
dark energy and background matter interact through an interaction
term $Q$. Thus the equations of the conservation of energy in
(\ref{a1}) and (\ref{a2}) should be changed into
 \begin{eqnarray}\label{b1}
 &&\dot{\rho}_y+3\frac{\dot{a}}{a}(\rho_y+p_y)=-Q,\label{doty}\\\label{b2}
 &&\dot{\rho}_m+3\frac{\dot{a}}{a}\rho_m=Q,\label{dotm}
 \end{eqnarray}
and the equation for radiation in (\ref{a3}) is still held. The sum of Eqs. (\ref{b1}) and (\ref{b2}) guarantees that the total energy of YMC and dust matter is still conserved. It is worth noting that the free Yang-Mills equation in (\ref{ym_equation}) is not satisfied when $Q\neq0$.
In the natural unit, the interaction term $Q$ has the dimension of $[{\rm energy}]^5$. The coupling $Q$ is phenomenological, and their specific forms will be addressed later. When $Q>0$, the YMC transfers energy into the matter, and this could be implemented, for instance, by the processes with the YMC decaying into pairs of matter particles. On the other hand, when $Q<0$, the matter transfers energy into the YMC. Note that, once $Q$ is introduced as above, it will bring another new parameters in the models, in addition the free parameter $\kappa$.

We introduce the following dimensionless variables, rescaled by the critical energy density $b\kappa^2/2$ of the YMC,
 \begin{eqnarray}
 x\equiv\frac{2\rho_m}{b\kappa^2},~~f\equiv\frac{2Q}{b\kappa^2H},
 \end{eqnarray}
where $f$ is the function of $x$ and $y$. By the help of the
definition of $y$, the evolution equations (\ref{doty}) and
(\ref{dotm}) can be rewritten as a dynamical system, i.e.
\begin{eqnarray}
 y'&=&-\frac{4y}{2+y}-\frac{f(x,y)}{(2+y)e^y},\label{y'}\\
 x'&=&-3x+f(x,y).\label{x'}
\end{eqnarray}
Here, a prime denotes derivative with respect to the so-called
e-folding time $N\equiv\ln a$. The fractional energy densities of
dark energy and background matter are given by
 \begin{eqnarray}
 \Omega_y=\frac{(1+y)e^y}{(1+y)e^y+x},~~{\rm
 and}~~\Omega_{m}=\frac{x}{(1+y)e^y+x}.\label{Omega}
 \end{eqnarray}

 Before discussing the specific form of the interaction term $Q$, let us first investigate the general feature of this dynamical system, described by (\ref{y'}) and (\ref{x'}).
We can obtain the critical point $(y_c,x_c)$ of the autonomous
system by imposing the conditions $y_c'=x_c'=0$. From the
equations (\ref{y'}) and (\ref{x'}), we obtain that the critical
state satisfies the following simple relations
 \begin{eqnarray}
 3x_c&=&f(x_c,y_c),\label{[1]}\\
 3x_c&=&-4y_ce^{y_c}.\label{[2]}
 \end{eqnarray}
So we can obtain the critical state $(y_c,x_c)$ by solving these
two equations. In order to study the stability of the critical
point, we substitute linear perturbations $y\rightarrow y_c+\delta
y$ and $x\rightarrow x_c+\delta x$ about the critical point into
dynamical system equations (\ref{y'}) and (\ref{x'}) and linearize
them. Thus, two independent evolutive equations are derived,
\[
\left(
 \begin{array}{c}
 \delta y'\\
 \delta x'
  \end{array}
 \right)
 \equiv
M \left(
 \begin{array}{c}
 \delta y\\
 \delta x
  \end{array}
 \right)  =
\left( \begin{array}{cc}
G_{,\rm y}+R_{,\rm y} & R_{,\rm x} \\
f_{,\rm y} & f_{,\rm x}-3
 \end{array}
 \right)
\left(
 \begin{array}{c}
 \delta y\\
 \delta x
  \end{array}
 \right),
\]
where
 $R_{,\rm y}\equiv\partial R/\partial
 y$ at  $(y,x)=(y_c,x_c)$. The definitions of $R_{,\rm x}$, $f_{,\rm y}$, $f_{,\rm x}$ and
$G_{,\rm y}$ are similar. The functions $G$ and $R$ are defined by
 \[
 G\equiv G(y)=-\frac{4y}{2+y},~~~
 R\equiv R(x,y)=-\frac{f(x,y)}{(2+y)e^y},
 \]
which are used for the simplification of the notation. The two
eigenvalues of the coefficient matrix $M$ determine the stability
of the corresponding critical point. The critical point is an
attractor solution, which is stable, only if both the these two
eigenvalues are negative (stable node), or real parts of these two
eigenvalues are negative and the determinant of the matrix $M$ is
negative (stable spiral).

Here we discuss some general features of the attractor solutions,
regardless the special form of the interaction term $Q$. From the
expression (\ref{[2]}), we find that $x_c=-(4/3)y_ce^{y_c}$.
Substitute this into the formula (\ref{Omega}), one obtains
 \begin{eqnarray}
 \Omega_y=\frac{(y_c+1)e^{y_c}}{(y_c+1)e^{y_c}+x_c}=\frac{3+3y_c}{3-y_c}.\label{Omegay}
 \end{eqnarray}
Since $0\leq\Omega_y\leq1$, this formula follows a constraint of
the critical point
 \begin{eqnarray}
 -1\leq y_c\leq0.\label{constrainty}
 \end{eqnarray}
From the formulae (\ref{eos}) and (\ref{Omegay}), we find a simple, but interesting relation,
 \begin{eqnarray}
 \Omega_yw_y=-1.\label{-1}
 \end{eqnarray}
This relation is held for all attractor solutions, independent of
the special form of the interaction. Since the value of $\Omega_y$
is smaller than or equal to unity in the attractor solution, we
obtain that
 \begin{eqnarray}
 w_y\leq-1.
 \end{eqnarray}
This means that, the EOS of the YMC dark energy must be smaller than or equal to $-1$, i.e.
phantom-like or $\Lambda$-like. Since in the early universe, the
value of the order parameter of the YM field $F$ is much larger
than that of $\kappa^2$, i.e. $y\gg1$, the YM field is a kind of
radiation component. However, in the late attractor
solution, the dark energy is phantom-like or $\Lambda$-like. So
the phantom divide must be crossed in the former case, which is different from
the interacting quintessence models.

It is interesting to investigate the total EOS of the YMC and dust matter, which determines the finial fate of the universe, when the radiation becomes negligible.
The total EOS is defined by
 \begin{eqnarray}
 w_{tot}\equiv\frac{p_{tot}}{\rho_{tot}}=\frac{p_y+p_m}{\rho_y+\rho_m}=\Omega_yw_y,
 \end{eqnarray}
where $p_m=0$ is used. From the relation (\ref{-1}), we obtain
that, in the attractor solution,
 \begin{eqnarray}
 w_{tot}=-1.
 \end{eqnarray}
This result is also independent of the specific form of the
interaction. So the universe is an exact de Sitter expansion, and
the cosmic big rip is naturally avoided, although the YMC
dark energy can be phantom-like.

~

Now, let us discuss the evolution of the various components in the universe for some specific interaction models. In this paper, we shall focus on the following three phenomenological
interaction models: $Q\propto H\rho_y$, $Q\propto H\rho_m$ and $Q\propto H(\rho_y+\rho_m)$, separately. Some other phenomenological models have also been discussed in the papers \cite{xzzcqg2007,zijmpd2009}.

\subsubsection{$Q\propto H\rho_y$}

In this case, we can write the function $f(x,y)$ as the following form, $f(x,y)=\alpha(1+y)e^y$. Of course, when $\alpha=0$. the system returns to the model with free YMC dark energy. Here, we consider the simplest case with $\alpha$ being a non-zero dimensionless
constant.  Thus, the dynamical equations in (\ref{y'}) and (\ref{x'}) becomes
 \begin{eqnarray}
  y'&=&-\frac{4y}{2+y}-\frac{\alpha(1+y)}{(2+y)},\label{y'1}\\
 x'&=&-3x+\alpha(1+y)e^y.\label{x'1}
 \end{eqnarray}
Obviously, the evolution of dust matter and YMC are influenced by the interaction by the function $f(x,y)$. We expect, when the fraction density of YMC was sub-dominant in the universe, the effect on the dust is small. Only in the latest stage of the universe, where the YMC dark energy dominates the evolution of the universe, the effect of interaction on the dust becomes important.

The critical point ($y_c,x_c$) is obtained by imposing the condition $y'_c=x'_c=0$, which are
 \begin{eqnarray}
 y_c=-\frac{\alpha}{4+\alpha},
 ~~x_c=-\frac{4y_c}{3}e^{y_c},
 \end{eqnarray}
and the fractional energy density and the EOS of the YMC at this critical point are
 \begin{eqnarray}
\Omega_y=-\frac{1}{w_y}=\frac{3}{\alpha+3}.
 \end{eqnarray}
The constraint $0\le\Omega_{y}\le1$ requires that
$\alpha>0$ (note, we have set $\alpha\neq0$ throughout the discussion). In this condition, we find that $w_y<-1$ is satisfied.
In order to keep this critical point being stable, i.e. ($y_c,x_c$) is the attractor solution, a constraint on $\alpha$ can be derived: $\alpha>-8$, which is auto-satisfied when $\alpha>0$ is required. So the attractor solution requires that the constraint on the coefficient being
\begin{eqnarray}\label{require1}
\alpha>0,
\end{eqnarray}
 which follows that the EOS of YMC in this attractor solution must be negative. If we still require that, the fraction density of YMC is larger than the value at the present stage, i.e. $\Omega_y>0.73$, the constraint on $\alpha$ becomes much tighter $0<\alpha<1.11$.

In order to have a much clear picture for this system, let us
study the evolution of the various components in the universe by
adopting $\alpha=0.5$. We still choose the initial condition at
the BBN stage, i.e. $z=10^{10}$, where the value of $x$ is chosen
as $x_i=1.8\times10^{-29}$ to keep the present value of $\Omega_y$
being $\Omega_y=0.73$. For the YMC, we consider the following two
choices as the initial condition: i.e. $y_i=60$ and $y_i=20$. In
the former case, we have $\Omega_{yi}/\Omega_{mi}=0.16$, and in
the latter case, we have $\Omega_{yi}/\Omega_{mi}=3\times
10^{-19}$. Although, the difference between these two
$\Omega_{yi}/\Omega_{mi}$ cases is larger than 17 orders. we will
show that, these two models follow the similar present state of
universe.

We should also fix the value of $\kappa$. In order to keep the present total energy density in the universe being $\rho_{tot}\simeq8.099h^2\times10^{-11}{\rm eV}^4$, we find the $\kappa=4.00h\times10^{-5}{\rm eV}^{2}$ in both initial condiitons. If the Hubble constant $h=0.72$ is adopted, we get $\kappa=2.88\times10^{-5}{\rm eV}^{2}$. Again, we find the value of $\kappa$ is much smaller than the typical energy scale in the particle physics.

\begin{figure}
 \centerline{\includegraphics[width=18cm,height=10cm]{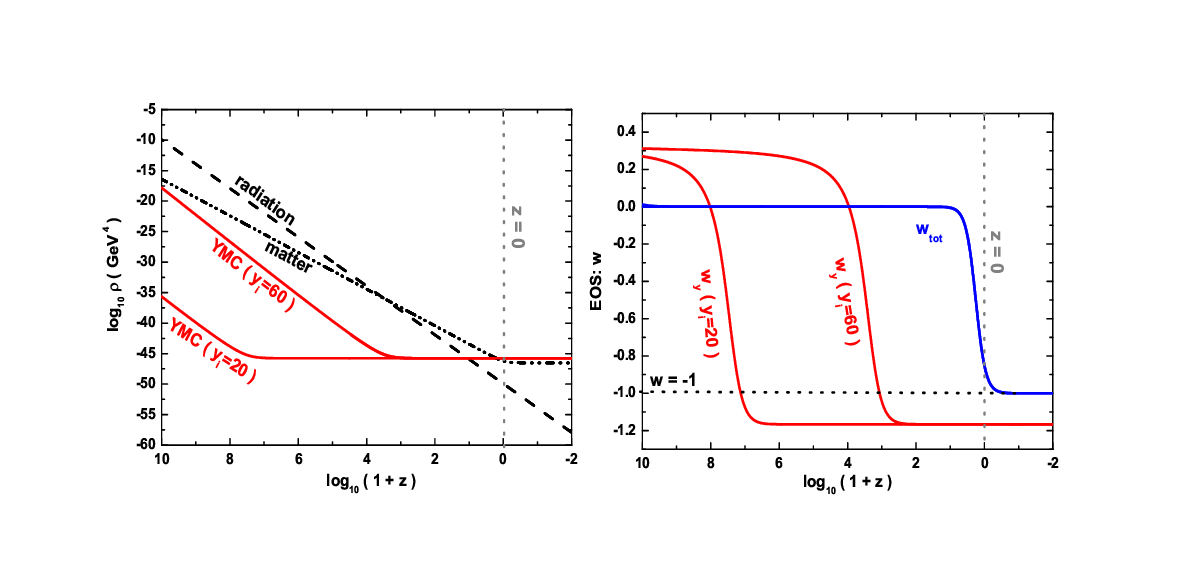}}
 \caption{In the coupled YMC dark energy models, the evolution of the YMC energy density (left panels) and EOS (right panels) for the models with different initial conditions.  \label{fig3}}
 \end{figure}

In Fig. \ref{fig3} (left panel), we plot the evolution of various components in the universe. Note that, the evolution line for dust component covered each other in both initial cases, as well as the radiation component. Similar to the free YMC models (shown in the left panel in Fig. \ref{fig1}), in the early stage of the universe, the YMC was tracking the radiation, and it transfered to the attractor solution in the later stage.   The smaller $y_i$ induces the earlier transfer.  In the right panel of Fig. \ref{fig3}, we plot the evolution of EOS of YMC $w_y$ and the total EOS $w_{tot}$. Note that, the evolution line $w_{tot}$ covered each other in both initial cases. As expected, in the both initial conditions, in the early stage, $w_y\rightarrow1/3$. In the late stage, $w_y$ runs to the attractor solution with $w_y=-(\alpha+3)/3$ with $\alpha=0.5$, i.e. $w_y\rightarrow-7/6$.  In the intermedial stage, the EOS $w_y$ must cross the line with $w_y=-1$. For the total EOS $w_{tot}$, in the early stage, $w_{tot}\rightarrow0$, where dust component is dominant than YMC. However, in the late stage, $w_{tot}\rightarrow-1$, which is consistent with the previous discussions.

\subsubsection{$Q\propto H\rho_m$}

Now, let us turn to the interaction cases with $Q\propto H\rho_m$, which is equivalent to the
form $f(x,y)=\beta x$. The dynamical equations in (\ref{y'}) and (\ref{x'}) becomes
\begin{eqnarray}
 y'&=&-\frac{4y}{2+y}-\frac{\beta x}{(2+y)e^y},\label{y'2}\\
 x'&=&(\beta-3)x.\label{x'2}
\end{eqnarray}
If we consider the simplest, where $\beta$ is a constant, the equation (\ref{x'2}) follows that $x\propto a^{\beta-3}$. From the definition of $x$, we derive that $\rho_m\propto a^{\beta-3}$. When $\beta=0$, the model returns to the free YMC cases, and $\rho_m\propto a^{-3}$ as usual. However, when $\beta\neq0$, the evolution of the dust component is changed, which is conflicted with the evolution of dust in the standard hot big-bang model. So it is dangerous to consider this kind of interaction term in the early stage of the universe.

However, it is allowed to consider the form of $Q\propto H\rho_m$ as a kind of phenomenological model in the late stage of the universe.  The critical point ($y_c,x_c$) of (\ref{y'2}) and (\ref{x'2}) is obtained by imposing the condition $y'_c=x'_c=0$, we find these is no solution at all. So we conclude that, it is impossible to obtain an attractor solution for this kind of system.

\subsubsection{$Q\propto H(\rho_y+\rho_m)$}

In the end, let us discuss the another kind of phenomenological model, where $Q\propto H(\rho_y+\rho_m)$ is satisfied. This is equivalent
to set the form $f(x,y)=\gamma[(y+1)e^y+x]$. We consider the simplest case, where $\gamma$ is a
non-zero dimensionless constant.  In the dark energy dominant stage, this system returns to the first case with $Q\propto H\rho_y$, and in the dust dominant stage, it returns to the second case with $Q\propto H\rho_m$. Similar to the second case, if we directly apply this system to the early universe, the evolution of dust would be changed, which is conflicted with the prediction of the standard hot big-bang models.

In this paper, we also consider this interaction form as a kind of phenomenological model in the late stage of the universe. The dynamical equations in (\ref{y'}) and (\ref{x'}) becomes
\begin{eqnarray}
 y'&=&-\frac{4y}{2+y}-\frac{\gamma[(y+1)e^y+x]}{(2+y)e^y},\label{y'3}\\
 x'&=&-3x+\gamma[(y+1)e^y+x].\label{x'3}
\end{eqnarray}
From the equations (\ref{[1]}) and
(\ref{[2]}), we obtain the critical point
 \begin{eqnarray}
 y_c=\frac{3\gamma}{\gamma-12},
 ~~x_c=-\frac{4y_c}{3}e^{y_c}.
 \end{eqnarray}
The fractional energy density and the EOS of the YMC at this critical point are
 \begin{eqnarray}
\Omega_y=-\frac{1}{w_y}=\frac{3-\gamma}{3}.
 \end{eqnarray}
 The constraint of $0\le\Omega_y\le1$ requires that $0<\gamma\leq3$.
In order to keep this critical point being stable, i.e. ($y_c,x_c$) is the attractor solution, another constraint of $\gamma$ can be derived: $\gamma>-120/31$, which is auto-satisfied when $0<\gamma\leq3$ is required. So the attractor solution requires that the constraint on the coefficient being
\begin{eqnarray}\label{require1}
0<\gamma\leq3,
\end{eqnarray}
 which follows that, the EOS of YMC in this attractor solution must be negative. If we still require that, the fraction density of YMC is larger than the value at the present stage, i.e. $\Omega_y>0.73$, the constraint on $\alpha$ becomes much tighter $0<\gamma<0.81$.


\section{Statefinder and $Om$ diagnosis in the YMC models}

 \begin{figure}
 \centerline{\includegraphics[width=18cm,height=10cm]{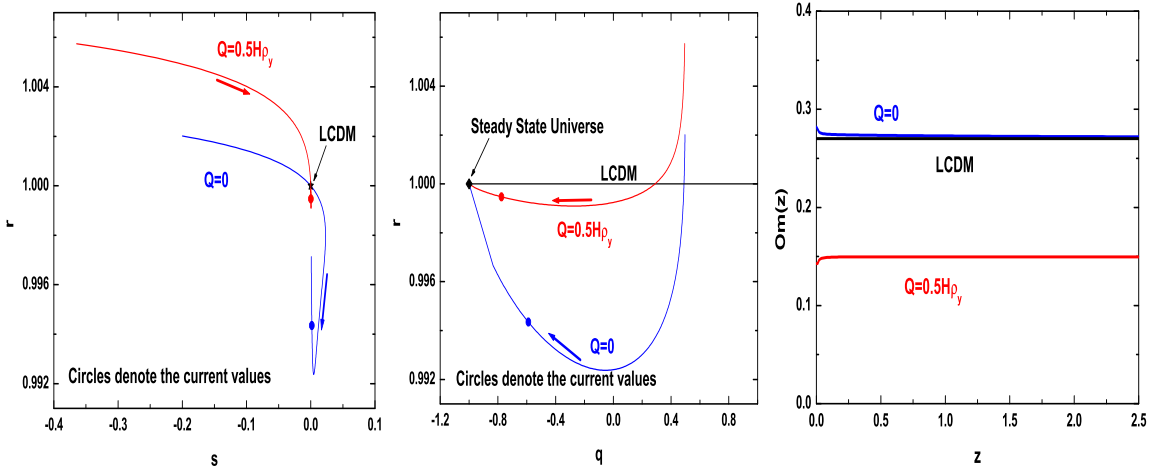}}
 \caption{Left panel: The $r-s$ diagram of the YMC models; Middle panel: The $r-q$ diagram of the YMC models; The right panel: $Om(z)$ diagnostic in the YMC models.\label{fig4}}
 \end{figure}

In this section, we shall present the way to discriminate between the YMC models and the other dark energy models. In the previous works \cite{statefinder}\cite{om}, the authors suggested to use the so-called ``statefinder" $\{r,s\}$ pair and $Om$ diagnostics. Here, we shall also apply the statefinder diagnosis into the YMC models. The similar discussion can be found in the previous works \cite{zijmpd2008}\cite{wzxjcap2008}\cite{tzxijmpd2009}\cite{tzprd2008}.

The statefinder diagnostic pair $\{r,s\}$ is defined as
\begin{equation}
r\equiv\frac{\dddot{a}}{aH^3},~~s\equiv\frac{r-1}{3(q-1/2)},
\end{equation}
where $q=-\ddot{a}/(aH^2)$ is the deceleration parameters. $r$ forms the next step in the hierarchy of geometrical cosmological parameters beyond $H$ and $q$, and $s$ is a linear combination of $r$ and $q$. Apparently, the statefinder parameters depends only on scale factor $a$ and its derivatives, and thus it is a geometrical diagnostic.  These parameters can also been expressed in terms of $\rho_{tot}$ and $p_{tot}$ as follows
\begin{equation}
q=\frac{1}{2}\left(1+\frac{3p_{tot}}{\rho_{tot}}\right),~~r=1+\frac{9(\rho_{tot}+p_{tot})\dot{p}_{tot}}{2\dot{\rho}_{tot}p_{tot}},~~s=\frac{\rho_{tot}+p_{tot}}{p_{tot}}\frac{\dot{p}_{tot}}{\dot{\rho}_{tot}}.
\end{equation}

By using Eqs. (\ref{b1}) and (\ref{b2}), we plot the trajectories
of $(r,s)$ and of $(r,q)$ in Fig. \ref{fig4} from the redshift
$z=10$, where we have considered two cases with $Q=0$ and
$Q=0.5H\rho_{y}$. The arrows along the curves indicate the
direction of evolution. In both cases, we have chosen the initial
condition at  the redshift $z=3454$, where the densities of matter
and radiation equate to each other. We choose the initial
condition to make the $\Omega_{y}=0.01$ at this high redshift.

From this figure, we find that, the model with different
interaction terms have the difference in both $(r,s)$ and $(r,q)$
trajectories. However, in the low redshift ($z<10$), especially
when $z\sim0$, the trajectories in both models become quite close
to those of the $\Lambda$CDM model. However, we should also
mention that, the trajectories of the statefinder shows in this
figure are quite different from the other dark energy models, such
as the decaying vacuum model, the Quintessence models, the
K-essence models (see for instant \cite{tzprd2008} ). In order to
break the degeneracy between the YMC models and the $\Lambda$CDM
models, let us consider the $Om$ diagnostic, which is defined as
\begin{equation}
Om(x)=\frac{h^2(x)-1}{x^3-1},
\end{equation}
where $x\equiv(1+z)$, and $h(x)\equiv H(x)/H_0$. Thus $Om$ involves only the first derivative of the scale factor through the Hubble parameter and is easier to reconstruct from the observational data. For the $\Lambda$CDM model, it is simple, i.e. $Om(x)=\Omega_m$, independent of the redshift. For YMC model, we plot $Om$ as a function of $z$ in the Fig.\ref{fig4} (right panel). It is interesting to find that, for either model, $Om$ is nearly a constant at the low redshift. For the model without interaction, as expected, $Om\simeq0.27$, very close to that of $\Lambda$CDM models. However, for the model with $Q=0.5H\rho_y$, $Om=0.15$, which is helpful to differentiate the coupled YMC models from the $\Lambda$CDM model.


\section{Conclusion}

In order to answer the observed accelerating expansion of the universe in the present stage, we introduce the quantum Yang-Mills condensate dark energy models.
Different from the general scalar field models, the quantum Yang-Mills fields are the indispensable cornerstone to particles, and
the Lagrangian of the Yang-Mills field is predicted by quantum corrections according to field theory.

In this paper, we review the main characters of the YMC dark energy models, where both free YMC model and possible coupled YMC models are considered. In all these models, the EOS of YMC is close to $w_y=1/3$ in the early stage, and tracked the evolution of radiation components. In the late stage, the value of $w_y$ runs to the attractor solution with $w_y=-1$ for the free YMC model, or with $w_y<-1$ for the coupled YMC models. This naturally explains the observations. The present state of the universe is independent of the initial state of Yang-Mills field, and the coincidence problem is naturally avoided. In the coupled YMC models, not only the state of $w_y<-1$ can be naturally realized, mildly suggested by observations, but also the `big-rip' problem is auto-avoided. The most importance we should mention is that, as a dynamic dark energy model motivated by quantum effective YM field Lagrangian, all the YM models based upon 1-loop, 2-loop, and 3-loop quantum corrections, respectively, have similar dynamic behaviors. That is, the main properties of YM model for dark energy remain stable when the number of loops for quantum corrections increases up to 3-loop.

However, in all the models, the fine-tunning problem still exists, i.e. the value of $\kappa\sim10^{-5}$eV$^{2}$, which is too low comparing with the typical scales in particle physics.  We should notice that, this problem exists in nearly all the dark energy models, which may suggest a new physics at this low energy scale.


\end{document}